%% file: neurips_2023.tex
\documentclass{article}

\PassOptionsToPackage{numbers, compress}{natbib}

\usepackage[preprint]{neurips_2023}

\usepackage{blindtext}
\usepackage{wrapfig}
\usepackage{multirow}
\usepackage{amsmath}
\usepackage[utf8]{inputenc} 
\usepackage[T1]{fontenc}    
\usepackage{hyperref}       
\usepackage{url}            
\usepackage{booktabs}       
\usepackage{amsfonts}       
\usepackage{nicefrac}       
\usepackage{microtype}      
\usepackage{xcolor}         
\usepackage[pdftex]{graphicx}
\usepackage{arydshln}
\usepackage{caption}
\newcommand{\METHODNAME}{MSA-Augmenter}
\newcommand{\REAL}{\textit{real-world challenging MSAs }}
\newcommand{\ART}{\textit{artificial challenging MSAs }}
\title{Enhancing the Protein Tertiary Structure Prediction by Multiple Sequence Alignment Generation}

\author{%
  Le Zhang$^{3,1}$\thanks{Equal Contribution. The works were completed during Le Zhang's internship at Shanghai AI Lab.},  Jiayang Chen$^{4*}$, Tao Shen$^5$, Yu Li$^{4\dag
}$, Siqi Sun$^{2,1\dag}$\\
  $^1$ Shanghai AI Laboratory \\
  $^2$ Fudan University \\
  $^3$ Mila, Université de Montréal \\ 
  $^4$ The Chinese University of Hong Kong\\
  $^5$ Zelixir Biotech\\
  \texttt{le.zhang@mila.quebec, liyu@cse.cuhk.edu.hk, siqisun@fudan.edu.cn}
}

\begin{document}

\maketitle

\begin{abstract}
  \input{content/abstract}
\end{abstract}

\section{Introduction}
  \input{content/introduction}
\section{Related Work}
   \input{content/related_work}
\section{\METHODNAME{}}
\label{method}
  \input{content/methods}

\section{Results}
   \input{content/results}

\section{Conclusion}
We introduce \METHODNAME{}, a transformer-based seq2seq model for homogeneous protein sequence generation, represented as multi-sequence alignment, in a zero-shot fashion. The model, trained on a group sequence generation task, enables efficient parallel generation of co-evolutionary sequences. Our experimental results demonstrate the model's robust generalization capability to unseen protein families, particularly in generating high-quality sequences valuable for protein folding tasks when MSAs are of low quality. Furthermore, we highlight that the current result filtering criteria employed in AlphaFold2 could be improved, underlining the potential of sequence generation in MSA-based tasks.

\section*{Limitations}
While \METHODNAME{} significantly improves protein structure prediction, it is not without limitations: (1) Our method generates data to bolster downstream tasks, leading to inherent variability in results. Consequently, conducting more trials tends to yield better outcomes. (2) Despite our unique approach to minimize MSA computation, managing excessively long MSAs remains a challenge due to the computational limitations of the Transformer model. (3) The method's applicability may be restricted as it may not handle certain protein sequences not included in the pre-training data accurately. Given the extensive size of current protein databases and our pre-training data of only 2 million instances, this limitation is foreseeable. However, following the example of large language models that exhibit rapid growth and emergent capabilities, we anticipate that scaling up both the model size and the dataset could yield improved results. (4) Given the constraints of our evaluation dataset, we present results exclusively from CASP14. However, as CASP14 is among the most prestigious benchmarks and has served as a testing ground for numerous
 significant studies, this limitation is deemed acceptable.

\bibliographystyle{plain}
\bibliography{custom}

\input{content/appendix}


\end{document}

%% file: content/abstract.tex
The field of protein folding research has been greatly advanced by deep learning methods, with AlphaFold2 (AF2) \citep{jumper2021highly} demonstrating exceptional performance and atomic-level precision. As co-evolution is integral to protein structure prediction, AF2's accuracy is significantly influenced by the depth of multiple sequence alignment (MSA), which requires extensive exploration of a large protein database for similar sequences. However, not all protein sequences possess abundant homologous families, and consequently, AF2's performance can degrade on such queries, at times failing to produce meaningful results. To address this, we introduce a novel generative language model, \METHODNAME{}, which leverages protein-specific attention mechanisms and large-scale MSAs to generate useful, novel protein sequences not currently found in databases. These sequences supplement shallow MSAs, enhancing the accuracy of structural property predictions. Our experiments on CASP14 demonstrate that \METHODNAME{} can generate \textit{de novo} sequences that retain co-evolutionary information from inferior MSAs, thereby improving protein structure prediction quality on top of strong AF2. We release our code at \url{https://github.com/Magiccircuit/MSA-Augmentor}

%% file: content/introduction.tex
\begin{figure}[t]
    \centering
    \includegraphics[width=0.5\textwidth]{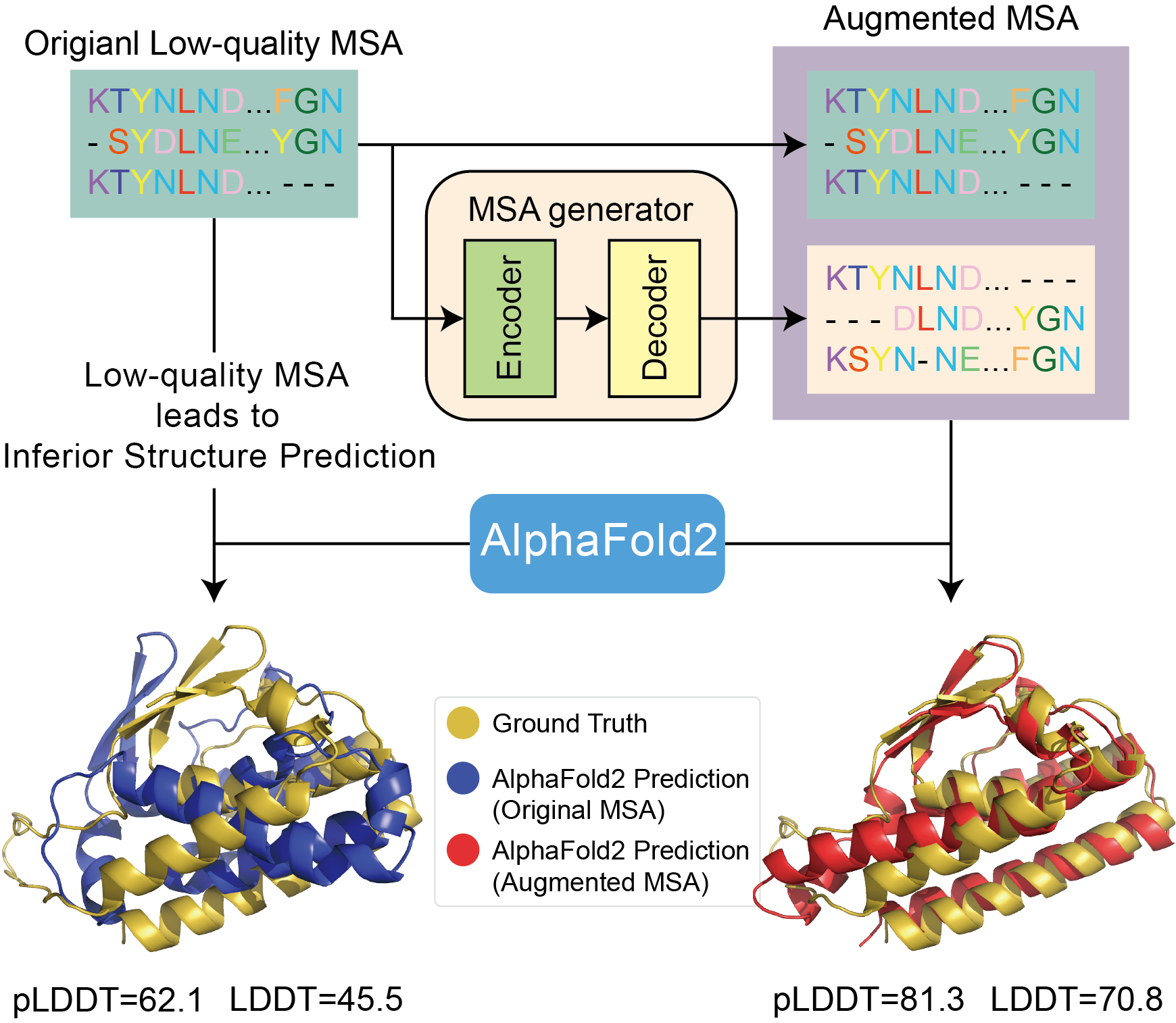}
    \captionsetup{font=small}
    \caption{\textbf{Procedure overview.} A low-quality MSA is introduced into the encoder as the source input, from which the decoder concurrently generates multiple homologous sequences. Subsequently, these sequences are appended to the low-quality MSA to create an enhanced MSA, which is then supplied to the downstream task. }
    \label{fig:overview}
\end{figure}

The rapid advancement of deep learning techniques has significantly addressed the longstanding challenge of protein folding prediction in structural biology. AlphaFold2 (AF2) \cite{jumper2021highly} has demonstrated remarkable precision in this task, largely attributed to its use of multiple sequence alignments (MSAs) — collections of diverse homologous sequences with same length that share evolutionary information. However, issues arise when few homologous sequences can be found for a target protein, leading to an inferior MSA. Performance of MSA-based models, including AF2, is observed to decline dramatically with such low-quality MSAs \cite{jumper2021highly, wang2022contact}. Moreover, constructing an MSA is labor-intensive and time-consuming due to the necessity of searching millions or billions of sequences in rapidly expanding protein databases.

The significant strides made in natural language processing (NLP), particularly in context-conditioned text generation, offer potential solutions to the challenges of protein research. Notably, the advent of transformer-based language models \citep{raffel2020exploring,touvron2023llama,flant5,codex} has demonstrated exceptional performance in various generation tasks, suggesting its applicability beyond text. By leveraging these advancements, we propose to enhance inferior MSAs by treating protein sequences akin to text, thus improving downstream tasks such as protein structure prediction.

To the best of our knowledge, no existing work on protein sequence generation incorporates both structural and evolutionary information from MSAs or aims to improve tertiary structure prediction accuracy. We introduce a novel method, \METHODNAME{}, a transformer-based seq-to-seq model designed for simultaneous generation of multiple sequences while considering global structural information from input MSAs. Our approach enables rapid, \textit{de novo} sequence generation, inferior MSA augmentation (figure \ref{fig:overview}), and adapts to protein domains, addressing computational challenges.

To summarize the main contribution of this article:
\begin{itemize}
    \item We present a novel generative transformer, designed to simultaneously generate multiple sequence alignments (MSAs) based on a provided MSA.
    \item We devise an effective ensemble strategy that aggregates results from various runs, delivering optimal outcomes.
    \item Our empirical findings affirm the efficacy of the generated MSAs as an additional source of evolutionary information for challenging protein structure prediction tasks. These results, when used in conjunction with the robust AF2, significantly surpass the existing state-of-the-art AF2 in the field
\end{itemize}

%% file: content/related_work.tex
\paragraph{Transformers} 

The Transformer model architecture, which is based on self-attention, has become popular in various fields like natural language processing, computer vision, and speech processing. Introduced as a sequence-to-sequence(seq2seq) model for machine translation\cite{vaswani2017attention}, Transformers aggregate source input information for decoder use in sequence generation. However, despite its wide application, its length quadratic complexity limits its use in handling sequences longer than 2000 tokens, which is especially problematic for protein MSAs, typically comprising long sequences.

Several architectures\citep{lewis2019bart,raffel2020exploring,fedus2021switch, flant5, ho2019axial,Beltagy2020LongformerTL,Zaheer2020BigBT,Katharopoulos2020TransformersAR,Dai2020FunnelTransformerFO} have been proposed to improve natural language understanding and generation. However, these methods cannot encode structural information within the standard Transformer architecture. Nevertheless, unsupervised learning approaches using Transformers in the protein sequence domain have been promising, for example, \cite{rives2021biological} demonstrate the potential of transformer on the biological domain, \cite{vig2020bertology} and \cite{rao2020transformer} show the direct correlation between learned protein representation and the contact map. Notably, the MSA Transformer\cite{pmlr-v139-rao21a} show great improvement for downstream tasks and ESM-2\cite{lin2022language} have made significant strides in predicting structure, function, and other protein properties directly from individual sequences. Besidesm The ProGen series, as presented in \cite{madani2020progen,nijkamp2022progen2}, offers detailed control and understanding of protein model development's data and model roles. Despite the existence of MSA-based Transformers, they remain limited as they are encoder-only or single-sequence-based decoder-only and cannot learn MSA's global information to generate diverse, co-evolutionary MSAs.

\paragraph{Protein Structure Prediction}
Proteins, despite their vast diversity, comprise only 20 distinct amino acids. The physical structure of these proteins, dictating their function and properties, is crucial to understanding life's fundamentals. Recent advancements, notably AF2 \cite{jumper2021highly}, have greatly progressed the field, yet significant challenges remain. AF2's success is attributed to its effective use of MSAs, as single-sequence-based prediction methods \cite{chowdhury2021single, lin2022language,chowdhury2022single,wu2022high} tend to underperform in comparison. We address the limitations of MSA-based models on protein sequences with few homologs, where prediction often deteriorates or fails entirely.

\paragraph{Inferior MSA Enhancement}
Several studies have aimed to integrate co-evolutionary knowledge into models. For instance, Guo et al. \cite{guo2020bagging} used a self-supervised approach to enhance Position-Specific Scoring Matrix (PSSM) for protein secondary structure prediction, while Wang et al. \cite{wang2021pssm, wang2022contact} employed knowledge distillation strategies. Similarly, Sgarbossa et al. \cite{sgarbossa2022generative} leveraged an MSA transformer in an iterative masking process to generate novel sequences, outperforming Potts models on smaller families. However, none of these prior works have managed to demonstrate their model's effectiveness by improving upon AF2's tertiary structure prediction.

In this study, we present \METHODNAME{}, a sequence-to-sequence generative Transformer uniquely designed to manage Multiple Sequence Alignments (MSAs) with unrolled lengths surpassing 10,000. This model captures both row and column information and generates insightful MSAs, thereby improving downstream protein structure prediction. 

%% file: content/methods.tex
\begin{figure*}[ht]
    \centering
    \includegraphics[width=1\textwidth]{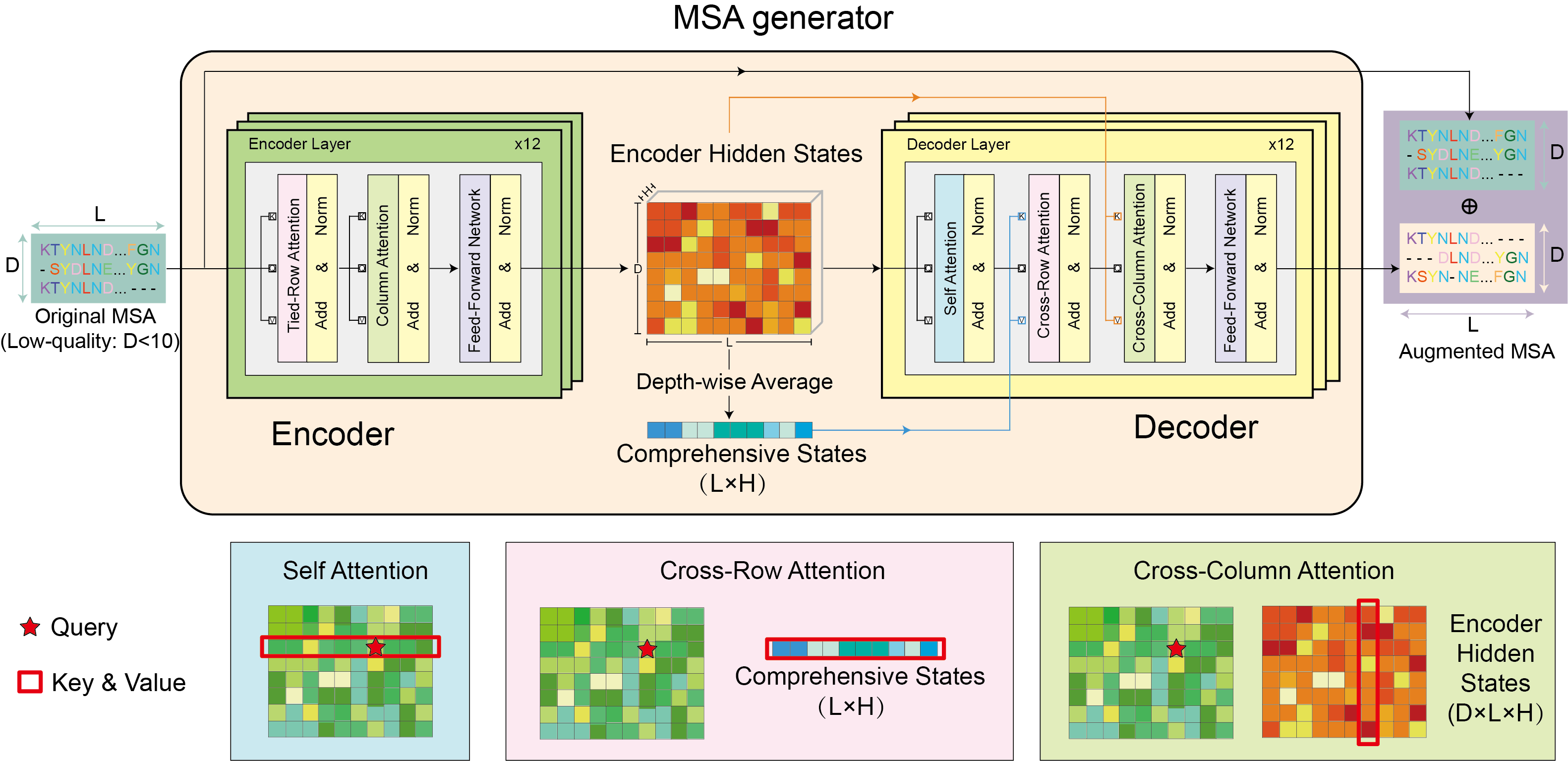}
    \caption{\textbf{\METHODNAME{} overview} (Top) Architecture and Pipeline and Module Attention Operations; (Bottom) Attention mechanism. In representing a single position, the red star denotes the query, while the red boxes signify keys and values for attention processing and computation.}
    \label{fig:method}
\end{figure*}

We introduce \METHODNAME{}, a model capable of accepting two-dimensional MSA input and generating an arbitrary number of co-evolutionary sequences in parallel. \METHODNAME{} is composed of a two-dimensional attention-based encoder and a decoder leveraging global information aggregation. The generation task is formalized in Sec. \ref{sec:task}, followed by detailed descriptions of the encoder and decoder design in Sec. \ref{sec:architecture}. The AlphaFold2 generation strategy is discussed in section \ref{sec:strategy}. Pertraining and model details are discussed in section \ref{sec:pretrain}

\subsection{Group Sequence Generation}\label{sec:task}


We introduce Group Sequence Generation (GSG), a novel task to create a cluster of homogeneous sequences from a single MSA input. Contrary to conventional machine translation tasks that map each target sequence to a single source, GSG generates multiple sequences concurrently, leveraging global evolutionary information (row-wise and column-wise) from a series of source sequences.

Our methodology simplifies the assimilation of source information by performing a depth-wise average of the encoder output. In the decoding phase, the decoder persistently accesses these comprehensive states for every decoding sequence, marking a departure from standard machine translation. The model's architecture and generation process are depicted in Figure \ref{fig:method}.

Our design's advantages are twofold. Firstly, it facilitates the seamless incorporation of global knowledge from the source input, which is crucial for generating homologous sequences. Secondly, our depth-wise average approach during the cross-row attention phase reduces computational complexity by shortening sequence length of unrolled MSA from $D \cdot L$ to $L$, where $D$ is input MSA depth and $L$ is sequence length of MSA. This strategy expedites decoding through parallel generation as opposed to auto-aggressive generation.

\subsection{\METHODNAME{} Architecture}\label{sec:architecture}

 \METHODNAME{} is a sequence-to-sequence transformer model that utilizes a bidirectional encoder to contextualize input MSA information and an auto-aggressive decoder to generate sequences based on this context. In order to adapt the input from text to MSAs, we employ the tied-row and column attention mechanism\cite{pmlr-v139-rao21a}, as shown in Figure \ref{fig:method}. This mechanism also reduces the computational complexity of modeling the input; for details, see \cite{pmlr-v139-rao21a}. Additionally, our model includes a cross-column module to process column-oriented information and a cross-row module to attend to global input context during decoding.

\paragraph{Tied-Row Attention}
Building on the shared structure of MSAs as discussed in \citep{pmlr-v139-rao21a}, we implement a specific attention mechanism that  that aggregates information by summing the attention map weights of rows from an input MSA $\in \mathbb{R}^{D \times L}$. For the $d^{th}$ row, the corresponding query, key, and value matrices are represented as $Q_{d}$, $K_{d}$, and $V_{d}$ respectively. The shared attention weight matrix is computed as:
\begin{equation}
    W_{TR}=softmax\left(\sum_{d=1}^D \frac{Q_d K_d^T}{\lambda(D,h)}\right) \in \mathbb{R}^{L \times L}
\end{equation}

Here, $\lambda(D,h)=\sqrt{Dh}$ is the square-root normalization to prevent the attention weights from linearly scaling with the number of sequences. The final output representation for the $d^{th}$ row of the MSA is given by $W_{TR}V_d$. In the decoder, we avoid tied-row attention to preserve diversity among the generated sequences and instead adhere to the original transformer self-attention as depicted in Figure \ref{fig:method}.

\paragraph{Column Attention} MSAs, with their columnar information, are a significant asset for protein sequence modeling, aiding tasks such as structure and secondary structure prediction. Incorporating the vertical-direction attention mechanism, originally introduced by \cite{ho2019axial}, we've integrated this mechanism into our \METHODNAME{} model.

Define $H^{col}i$ as the $i^{th}$ column of the hidden states $H \in \mathbb{R}^{D \times L \times h}$, with corresponding column query, key, and value matrices represented as $Q{i}$, $K_{i}$, and $V_{i}$. The attention weight matrix for column $i$ is calculated as:
\begin{equation}
    W_i=softmax(\frac{Q_i K_i^T}{\sqrt{h}})
\end{equation} Thus the final output representation for the $i^{th}$ column is thus $W_iV_i$. The cross-column attention operation remains identical, save for the query and key matrices being computed from the encoder hidden states as shown in Figure\ref{fig:method}.

\paragraph{Cross-Row Attention}
In sequence-to-sequence tasks like machine translation, attending to the input is essential as the model requires global information from the input MSA. Due to the computational complexity arising from excessive length, expanding the encoder hidden states into a single row for an MSA is computationally inefficient. To effectively harness the knowledge encoded in the encoder hidden states, we take the depth-wise average of the encoder output $H \in \mathbb{R}^{D \times L \times h}$ to form comprehensive states $H_{cs}=\frac{1}{D}\sum_{d=1}^DH_d \in \mathbb{R}^{L \times h}$. These states are used for cross-attention during decoding. With $K_d=H_{cs}W_k$ and $V_d=H_{cs}W_v$ representing the key and value matrices derived from the comprehensive states, and $Q_d$ denoting the query matrix from the $d^{th}$ row of decoder hidden states, we compute the weight matrix as follows: 
\begin{equation}
    W_{d}=softmax( \frac{ Q_d K_d^T}{\sqrt{h}})
\end{equation}
The final output representation is given by $W_dV_d$. As every decoding sequence attends to the same $K_d,V_d$, we can concurrently generate multiple sequences that share co-evolutionary knowledge with the input MSA.

\subsection{Pre-training objective}
We adapt the causal language modeling objective to Group Sequence Generation tasks. The loss for a source MSA $x \in \mathbb{R}^{D \times L}$ and target MSA $y \in \mathbb{R}^{D \times L}$ is defined as follows:
\begin{equation}
\mathcal{L}_{GSG}=\sum_{d=0}^D\sum_{l=0}^L\log(y^d_{l}|y^d_{<l},x)
\end{equation}
The probabilities are the output of the \METHODNAME{}, softmax normalized over the amino acid vocabulary independently per position $l$ in each sequence $d$.

\subsection{Generation and Ensemble Strategy}\label{sec:strategy}  
In order to encourage the generation of novel sequences and avoid repetition, we adopt a nucleus sampling approach with parameters \textit{top-p=50} and \textit{top-k=10}. Given \METHODNAME{}'s ability to simultaneously generate multiple sequences, it allows for potentially vast sequence generation. Utilizing this capability, we generate $n$ supplementary sequences for each MSA $\mathcal{M}$ with depth $m$. These sequences are then concatenated with the original input to create an augmented MSA $\mathcal {\widetilde{M}}$ with depth $m+n$.

Despite the ability to generate numerous sequences, we observed variability and inconsistency in their quality. To filter for high-quality sequences, we employ the pLDDT score, an AlphaFold2 confidence metric indicating the predicted outcome's quality, with higher scores suggesting more accurate predictions. By using the pLDDT score as a selection criterion, we augment each MSA multiple times as multiple trials, obtaining corresponding pLDDT scores for each trials. The MSA with the highest pLDDT score is then chosen as the ensemble optimal augmented outcome, and used to compute the LDDT score against the ground truth.

\subsection{Pretraining }
\label{sec:pretrain} 
\paragraph{Pretraining Dataset}
We have compiled a protein sequence dataset containing 2 million protein sequences, primarily featuring high-quality MSAs with a depth of up to 1000. The dataset was constructed in the following manner: Initially, we chose sequences from UniRef50 \citep{suzek2007uniref} as our queries. We then utilized the JackHMMER algorithm \citep{johnson2010hidden} to iteratively search for homologous sequences in UniClust30 \citep{mirdita2017uniclust} for the query sequences, repeating the process until no new sequences were found.

Each training instance comprises 30 sequences selected from an MSA, limited by the constraints of low-quality MSA input and GPU memory. We randomly select $m$ sequences ($m \in [2,10]$) as the "source context" MSA, with the remaining $30-m$ sequences serving as the "target" MSA. The choice of a small $m$ simulates challenging targets with few homologs in the MSA. Given that an MSA can be quite deep (>1000), our method of sampling a small subset allows for the creation of numerous low-quality MSAs from high-quality ones, which proves advantageous during multi-epoch algorithm training.
\paragraph{Models and Training details}
The \METHODNAME{} consists of 12 transformer encoders and decoders with 260M parameters where the embedding size is 768 and the number of heads is 12. The model is trained using the ADAM-W optimizer with a learning rate of $5e^{-5}$, a linear warm-up ratio of 0.01, and a square root decay scheduler. Due to the memory limitations, we set the batch size to 8. Eventually, the model is trained for 200k steps using 8 A100 GPUs.

%% file: content/results.tex
\subsection{Setup}

\paragraph{Evaluation Settings }\label{sec:eval set} 
Our research utilizes the same evaluation dataset, CASP14, as AF2, a highly respected dataset comprised of proteins from diverse biological families. Given the prohibitive cost of generating a large-scale protein structure prediction dataset, and considering that AF2 utilized all previously accessible structures for training, this dataset is the optimal choice for evaluation.

To assess \METHODNAME{}'s improvements, we assemble a test set from sequences in the CASP14. The MSAs of these sequences are constructed using JackHMMER, which searches the extensive UniRef90 database of 70 million sequences. We classify the CASP14 data into two sets according to MSA depth. The first set, our \REAL dataset, contains 11 sequences with an MSA depth of less than 10, posing substantial challenges to AF2's performance. The second set, the \ART dataset, consists of the remaining 81 targets with an MSA depth greater than 10, providing additional validation through an artificial challenge. All evaluations are carried out in a zero-shot manner, excluding all CASP14 sequences from the pre-training dataset.

\paragraph{Evaluation Metric} 
Local Distance Difference Tests (LDDT) \citep{mariani2013lddt} are widely used, superposition-free scores that assess local distance differences between actual and predicted structures. They range from 0 to 100, where 100 signifies a perfect match. To evaluate if \METHODNAME{} enhances the folding algorithm, we examine the LDDT differences in AlphaFold2's (AF2) output with low-quality MSAs before and after our augmentation.

For each MSA, AF2 generates five independent predicted structures along with corresponding predicted-LDDTs (pLDDTs), which gauge AF2's confidence in its predictions. Given the strong correlation between pLDDT and true LDDT, it serves as an essential criterion for assessing MSA quality. The prediction with the highest pLDDT is chosen as the final output \citep{jumper2021highly}.

\paragraph{Notations}\label{sec:notation} For clarity, our experiments utilize the following notations: `Org' refers to the original low-quality MSA results from AF2. `Aug-$n$' denotes the results from the augmented MSA for trial-$n$. Lastly, `Aug-EN' signifies the ensemble optimal MSAs sourced from multiple augmented MSAs, which represent the highest possible performance.

\begin{figure}[!htb]
    \centering
\includegraphics[width=0.8\textwidth]{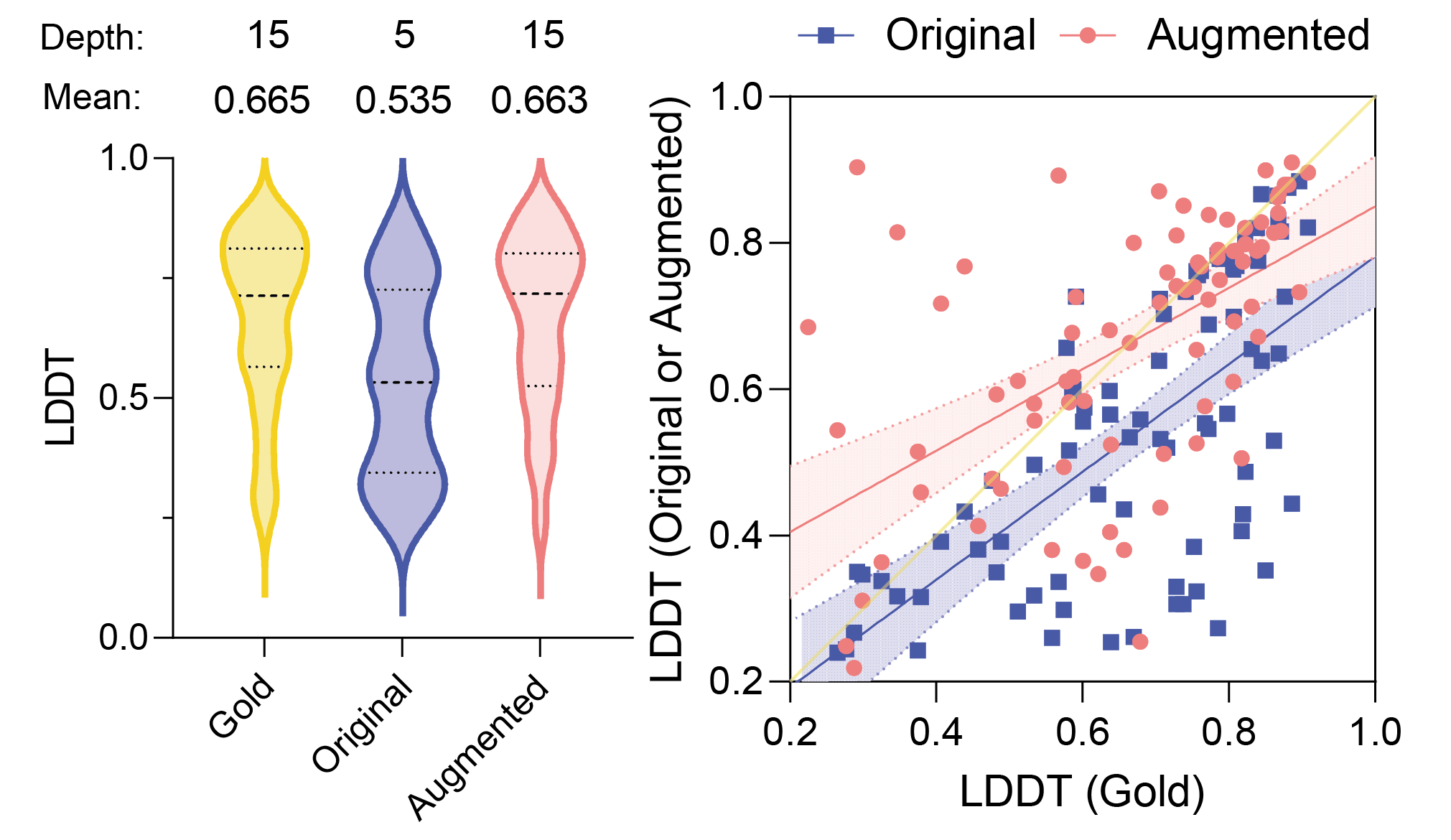}
    \caption{\textbf{(Left) LDDT distributions of AF2 predictions} across three distinct MSA categories within the artificial challenging cases. Quantiles are represented by dotted lines within the violin plots. \textbf{(Right) LDDT improvement of artificial challenging cases.} The x-axis represents LDDT inferred from the original MSA, while the y-axis represents LDDTs inferred from sampled and augmented MSA. }
    \label{fig:Qualitative_figures}
\end{figure}

\subsection{Artificial Challenging Cases} 

For a thorough evaluation, we selected 15 homologs from corresponding MSAs for all targets in our \ART dataset. This produced 81 MSAs of depth 15, 
\begin{wraptable}{r}{8cm}
\centering
\resizebox{7.5cm}{!}{
\begin{tabular}{ccccc}
\toprule
\multirow{2}{*}{$\Delta $LDDT} & \multicolumn{2}{c}{Augmented over Original} & \multicolumn{2}{c}{Augmented over Gold} \\ 
\cmidrule(lr){2-3}\cmidrule(r){4-5}
 & count & percentage & count & percentage \\\hline 
\textgreater{}40 & 11            & 13.6\%                         & 3                             & 3.7\%                          \\
(30,40{]}        & 7             & 8.6\%                          & 3                             & 3.7\%                          \\
(20,30{]}        & 10            & 12.3\%                         & 1                             & 1.2\%                          \\
(10,20{]}        & 8             & 9.9\%                          & 6                             & 7.4\%                          \\
(0,10{]}         & 23            & 28.4\%                         & 25                            & 30.9\%                         \\
\textless{}0     & 22            & 27.2\%                         & 43                            & 53.1\%                        \\\bottomrule
\end{tabular}
}
\captionsetup{font=small}
\caption{The distribution of LDDT improvements for the augmented MSA over the "original" and "gold standard" under the artificial challenging setting, where "Original" represents the down-sampled MSA with 5 sequences, and "Gold" represents the MSA with 15 sequences.}

\label{tab:improved_lddt_dist}
 \vspace{-3mm} 
\end{wraptable}transforming them into difficult-to-predict targets for AF2 by 
reducing homologs. These down-sampled MSAs, labeled `artificial gold', act as an upper-performance limit for our proposed approach. We further down-sampled these MSAs to five homologs, creating challenging targets (`artificial original') for enrichment by our model. We augmented these MSAs to a depth of 15 using \METHODNAME{} over three trials and then ensembled them (`artificial augmented') for comparison. As shown in figure \ref{fig:Qualitative_figures}(Left), the `artificial original' LDDT distribution significantly deteriorates compared to the `artificial gold' distribution, a gap that \METHODNAME{} can bridge by incorporating additional co-evolutionary information.

Figure \ref{fig:Qualitative_figures}(Right) demonstrates the overall enhancement in LDDTs for the \ART MSAs. Most of the `artificial original' LDDTs fall below the diagonal, while the majority of `artificial augmented' LDDTs are situated above, some substantially surpassing the `artificial original' LDDTs. \METHODNAME{} markedly enhances the overall structural prediction accuracy, achieving an average LDDT gain of \textbf{12.87} over `artificial original' MSAs. Details are presented in Appendix.

Table \ref{tab:improved_lddt_dist} presents the distribution of LDDT improvement \ref{tab:improved_lddt_dist}
highlighting that 11 sequences exhibit an enhancement greater than 40 LDDT – a substantial success for data augmentation. The most notable improvements stem from T1045s1-D1 (\textbf{+55.55}), T1070-D2 (\textbf{+55.34}), and T1068-D1 (\textbf{+54.72}), where AF2 struggles without augmentation. Interestingly, some augmented MSAs, such as T1032-D1 (\textbf{+46.02} LDDT), T1054-D1 (\textbf{+46.8} LDDT), and T1070-D2 (\textbf{+61.22} LDDT), outperform their `artificial gold' counterparts, highlighting the potential of our approach. However, 22 out of 81 MSAs underperform after augmentation, possibly due to the limited size of the pre-training dataset for structure prediction. As the dataset expands to include more co-evolutionary information, the effectiveness of the augmentation is expected to improve."

\subsection{Real-world Challenging Cases } \label{sec:real}

\begin{table*}[t]
\resizebox{1\textwidth}{!}{
\begin{tabular}{cccccccccccccc}
\toprule
\multirow{2}{*}{\textbf{ID}} &\multirow{2}{*}{\textbf{Depth}} & \multicolumn{6}{c}{\textbf{pLDDT}} & \multicolumn{6}{c}{\textbf{LDDT}} \\ 
\cmidrule(lr){3-8}\cmidrule(r){9-14}

                  &                  & Org & Aug1 & Aug2 & Aug3 & Aug-EN        & $\Delta $pLDDT & Org & Aug1 & Aug2 & Aug3 & Aug-EN         & $\Delta $LDDT  \\ \hline
\textbf{T1037-D1} & 4                & 36.0       & 40.7 & 36.0       & 36.0       & 40.7       & 4.4        & 24.1       & 26.0       & 24.1       & 24.1       & 26.1       & 2.0       \\
\textbf{T1042-D1} & 2                & 42.0       & 42.3 & 45.5       & 47.9       & 47.9       & 5.9        & 32.4       & 32.2       & 32.4       & 32.8       & 32.8       & 0.4        \\
\textbf{T1064-D1} & 9                & 60.3       & 66.9 & 63.1        & 65.0          & 66.9       & 6.6        & 31.3       & 37.5       & 31.3       & 51.4       & 37.5       & 6.2        \\
\textbf{T1074-D1} & 9                & 86.0       & 86.0 & 86.0       & 86.0       & 86.0       & 0.0           & 81.1       & 81.1       & 81.1       & 81.1       & 81.1       & 0.0           \\
\textbf{T1093-D1} & 3                & 62.1        & 81.3 & 62.1        & 62.1        & 81.3       & 19.2       & 45.5        & 70.8       & 45.5        & 45.5        & 70.8       & 25.3       \\
\textbf{T1093-D3} & 2                & 74.1       & 74.1 & 74.1       & 74.1       & 74.1       & 0.0        & 66.5        & 66.5        & 66.5        & 66.5        & 66.5        & 0.0          \\
\textbf{T1094-D2} & 7                & 90.5       & 90.5 & 93.3       & 91.5       & 93.3       & 2.8         & 77.1       & 77.1       & 77.0       & 76.5       & 77.0       & -0.1        \\
\textbf{T1096-D1} & 7                & 70.3       & 73.1 & 86.3       & 83.3       & 86.3       & 16.0       & 61.9       & 62.8       & 71.2       & 69.2       & 71.2       & 9.3        \\
\textbf{T1096-D2} & 2                & 45.6       & 50.0 & 54.6       & 50.2       & 54.6       & 9.0           & 34.1       & 36.6       & 37.0       & 34.7       & 37.0       & 2.9        \\
\textbf{T1099-D1} & 8                & 89.3       & 90.0 & 89.3       & 90.4       & 90.4       & 1.1        & 75.1       & 74.3       & 75.1       & 74.6       & 74.6       & -0.5        \\
\textbf{T1100-D2} & 2                & 42.2       & 47.3 & 42.2       & 42.2       & 47.3       & 5.1        & 34.9       & 35.9       & 34.9       & 34.9       & 35.9       & 0.9        \\ \hline
\textbf{Average}               & 5                 & 63.46 & 67.46 & 66.6 & 66.2 & 69.9 & 6.4 & 51.3 & 54.6 & 52.4 & 53.8 & 55.8 & 4.2 \\ \bottomrule
\end{tabular}}
\caption{pLDDT (left) and LDDT (right) improvement over challenging MSAs(depth<10).}
 \vspace{-5mm} 
\label{tab:huizong}
\end{table*}

Our ultimate objective with \METHODNAME{} is to enhance protein structure prediction in real-world, low-quality MSAs. We tested this approach on our constructed \REAL \ref{sec:eval set}).

Table \ref{tab:huizong} (left) shows improvements in pLDDT terms, demonstrating \METHODNAME{}'s efficacy as the average pLDDT increased by 6.39 in the challenging set. Further, multiple trials generally yielded better results due to the model's ability to generate valid sequences. Notably, significant improvements in pLDDT were observed for some sequences (e.g., T1093-D1, T1096-D1). However, some sequences, like T1074-D1 and T1093-D3, showed no pLDDT increase, suggesting that these sequences and their corresponding families might not be present in our pre-training dataset, thereby limiting the model's zero-shot generation capabilities.

While pLDDT is correlated with LDDT, it does not directly compare predicted structures to those derived from laboratory experiments. Consequently, it may be more practical to employ pLDDT as a selection criterion amongst a set of random augmentations. Following the ensemble strategy from Sec. \ref{sec:strategy}, we opted for the MSA with the highest pLDDT for LDDT computation with the ground-truth structures.

Table \ref{tab:huizong} (right) reveals that augmentation effectively improves structure prediction for low-quality MSAs. Specifically, 8 out of 11 MSAs were effectively augmented, leading to an average LDDT improvement of 4.2. The most considerable LDDT improvement was observed in T1093-D1 (with only three homologs), from 45.5 to 70.77. However, in certain MSAs, such as T1094-D2 (pLDDT +2.8, LDDT -0.1) and T1099-D1 (pLDDT +1.11, LDDT -0.5), an increase in pLDDT corresponded with a decrease in LDDT, suggesting that pLDDT may not always serve as a reliable selection criterion.

Although this study does not primarily aim for state-of-the-art performance, but to validate generative augmentation approaches for protein folding issues, there is potential for further improvement by optimizing hyper-parameters or modifying sampling strategies.

\subsection{How good is pLDDT as criteria}\label{sec:criteria}

\begin{wrapfigure}{r}{0.5\textwidth}  
  \centering
  \includegraphics[width=0.5\textwidth]{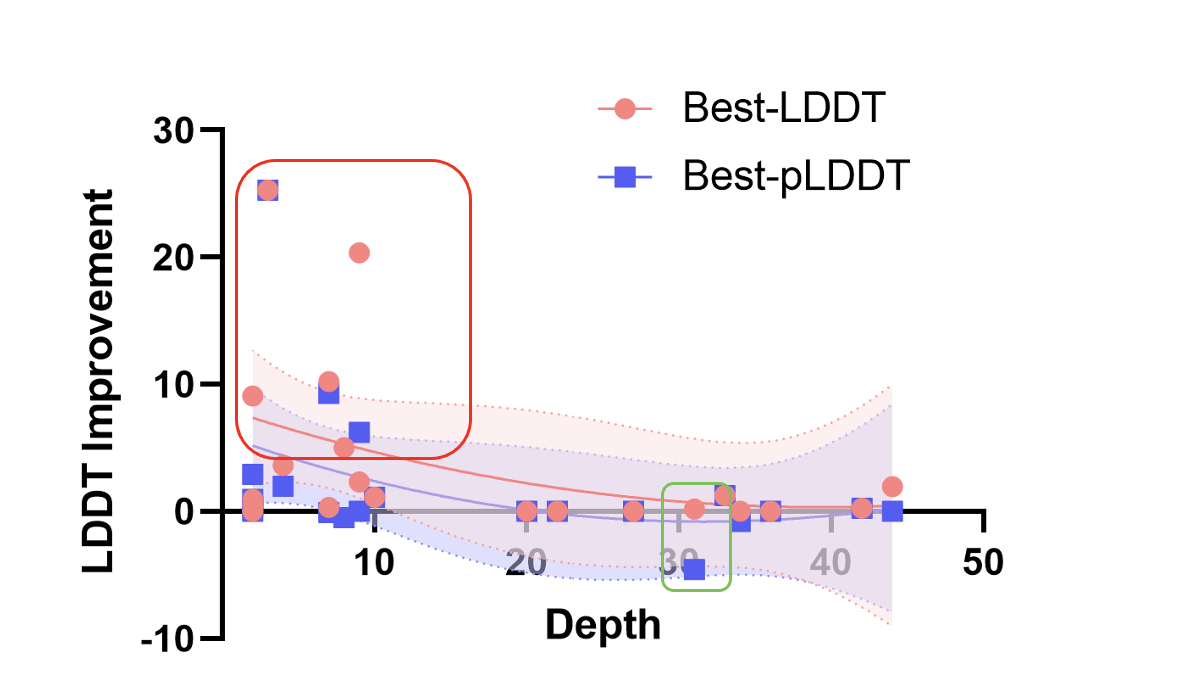}
    \captionsetup{font=small}   
   \caption{LDDT improvement on challenging MSAs, blue points representing MSA selected by ancillary pLDDT score and red points representing idealized MSA selected by LDDT}
    \label{fig:lDDT_plDDT}
     \vspace{-5mm} 
\end{wrapfigure}

To exhibit \METHODNAME{}'s capability independent of selection criteria, as pLDDT might not be the most suitable for selecting the final output, we conduct an idealized experiment. We select the highest LDDT score across all augmented MSAs, which involves calculating the LDDT with ground truth for each enhanced MSA, without using pLDDT as an auxiliary criterion. We use real-world MSAs from CASP14 with a depth of less than 50 as our test set to illustrate the trend. Detailed results for each MSA can be found in the Appendix, while the trending results are depicted in figure \ref{fig:lDDT_plDDT}

The substantial disparity between predictions selected by pLDDT and LDDT implies that pLDDT may not be the ideal criterion. For example, T1064-D1 and T1096-D2 show gaps of 14.12 and 6.17 respectively (indicated by the red rectangle), while T1099-D1 exhibits decreased performance (highlighted in the green rectangle). These examples suggest that our method has substantial untapped potential, contingent on improved selection criteria.

Notably, \METHODNAME{}'s performance varies greatly across MSAs of differing depth intervals. It doesn't appear effective for MSAs with moderate depth (10<depth<50), but significantly benefits the most challenging MSAs (depth<10). This could be due to two reasons: (1) The pre-training task mirrors this challenging setting, with input MSAs having depths between 2 and 10. Consequently, the model might struggle to encode sufficient information when given deeper MSAs by simply averaging the hidden states depth-wise. (2) As per AF2, prediction accuracy drops steeply for MSAs with depth less than 20, thus our method may not perform as effectively when augmenting MSAs with a depth greater than 20

\subsection{Case Study}
\begin{figure*}[!htb]
    \centering
    \includegraphics[width=0.9\textwidth]{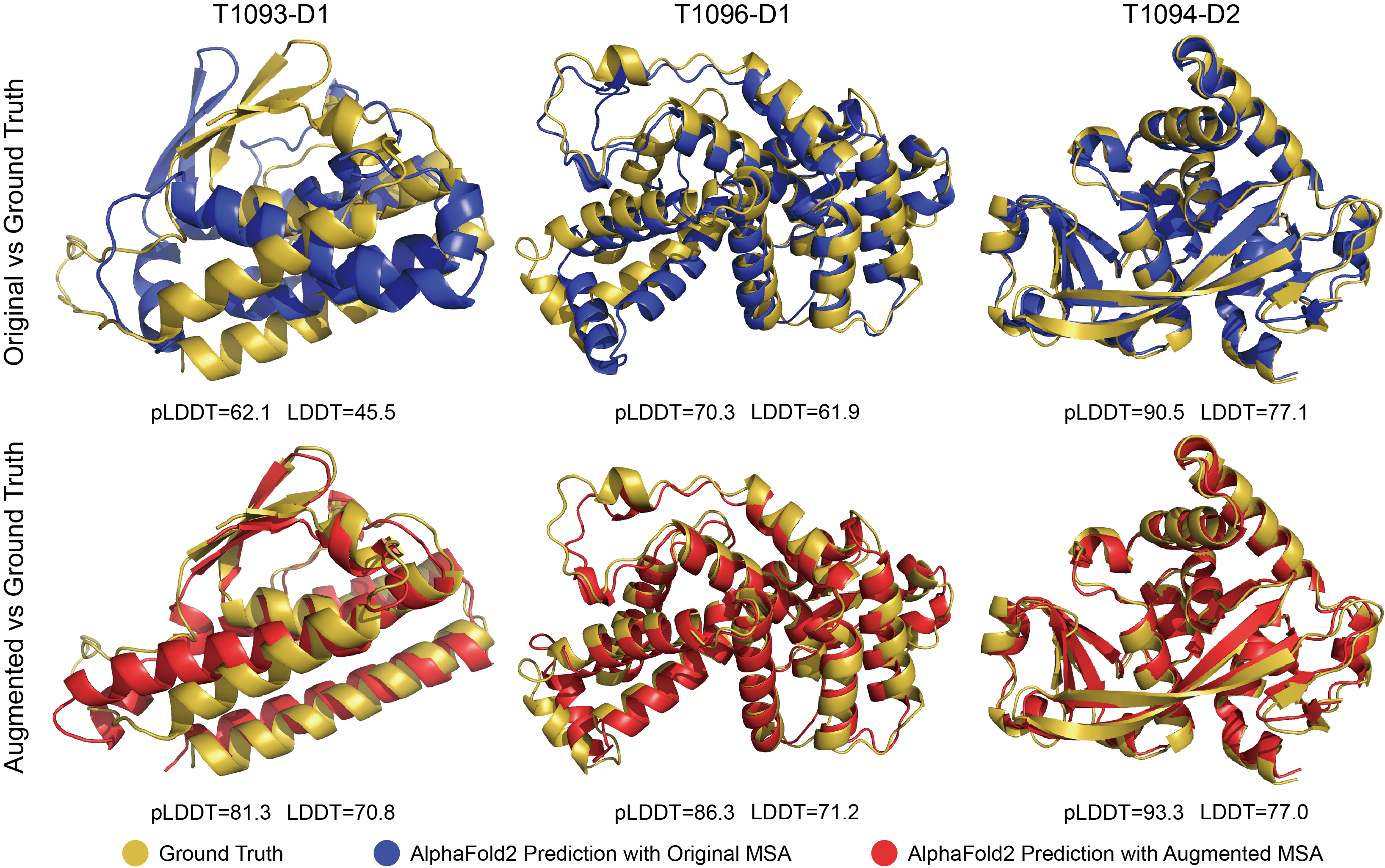}
    \caption{Protein structure visualization}
    \label{fig:protein_visualization}
\end{figure*}
We vividly demonstrate the efficacy of our method in enhancing MSA and aiding protein structure prediction tasks through visual results. We choose three MSAs (T1093-D1, T1096-D1, and T1094-D2) with respective depths of 2, 7, and 7, and corresponding LDDT improvements of 25.27, 9.27, and -0.1. We use PyMOL to visualize the protein structures, and the results are displayed in Fig \ref{fig:protein_visualization}. The corresponding MSAs are also provided in Appendix.


%% file: content/appendix.tex
\clearpage
\renewcommand{\thesubsection}{\Alph{subsection}}
\section*{Appendix}
\subsection{MSA Visualization}\label{sec:msa_visualizaiton}
\begin{figure*}[ht]
    \centering
    \includegraphics[width=1.0\textwidth]{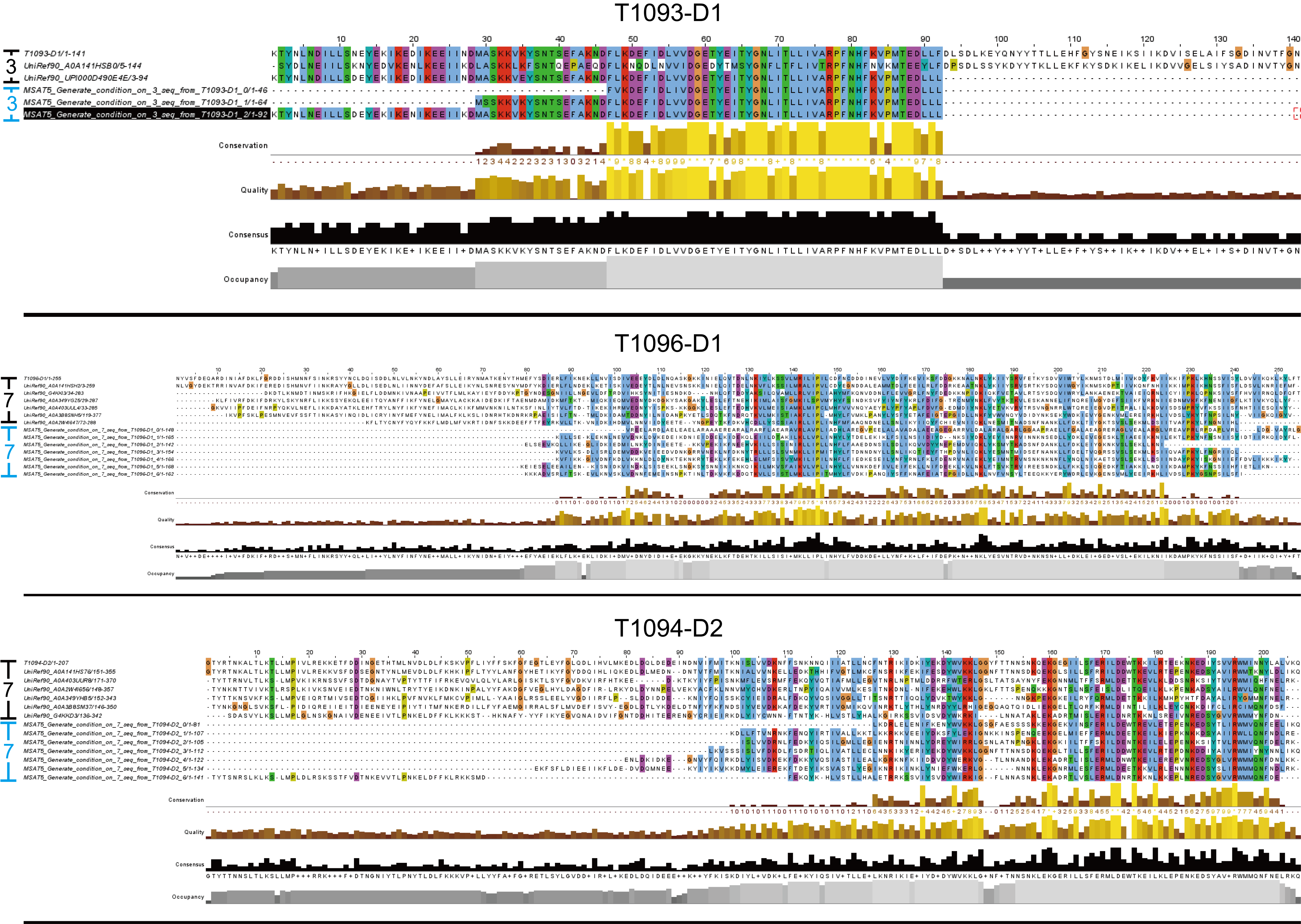} 
    \caption{colored-distribution MSA, different colors represent different amino acids in protein sequence, from top to bottom is T1060s2-D1, T1093-D1, T1096-D1. since \METHODNAME{} augment one times more sequences, the top half of each diagram represents original MSA, and bottom half represent generated MSA} 
    \label{fig:distribution}
\end{figure*}

We aim to investigate how the sequences in MSA vary with \METHODNAME{}, so we present MSA colored distribution in Fig \ref{fig:distribution}. We can see from the distribution in the column direction that the generated MSA shares similarities with the original ones but also has some variants that add diversity and encode external knowledge from the \METHODNAME{}. Second, we see long consecutive dashes in generated sequences, which is not surprising given that the dash is used for placeholder and alignment in multiple sequence alignment processes, and thus there are consecutive dashes in MSA sequences, the model learns this pattern and tends to generate continuous dash.

\subsection{Detailed Results}

\subsubsection{Real-Word Difficult and Challenging Results}\label{sec:all_results}
Results for Section \ref{sec:criteria} are displayed in Table \ref{tab:all_results} in detail and the comparison of different criteria for extremely challenging MSA (depth<10) is shown in table \ref{tab:criteria_detail}. We learn from the result that the \METHODNAME{} is capable of generating real high-quality MSAs but can't be selected under the current pLDDT criteria.

\begin{table}[h]
\centering
\resizebox{0.48\textwidth}{!}{
\begin{tabular}{cccccc}
       \toprule
      MSA-ID & Depth & Org  & by pLDDT & by LDDT\\ \hline
\textbf{T1037-D1}     & 4                          & 24.09                        & 1.99                 & 3.62                     \\
\textbf{T1042-D1}     & 2                          & 32.42                        & 0.42                 & 0.48                     \\
\textbf{T1064-D1}     & 9                          & 31.25                        & 6.23                 & 20.35                    \\
\textbf{T1074-D1}     & 9                          & 81.14                        & 0                    & 2.3                      \\
\textbf{T1093-D1}     & 3                          & 45.5                         & 25.27                & 25.27                    \\
\textbf{T1093-D3}     & 2                          & 66.5                         & 0                    & 0                        \\
\textbf{T1094-D2}     & 7                          & 77.12                        & -0.1                 & 0.31                     \\
\textbf{T1096-D1}     & 7                          & 61.92                        & 9.27                 & 10.23                    \\
\textbf{T1096-D2}     & 2                          & 34.07                        & 2.91                 & 9.08                     \\
\textbf{T1099-D1}     & 8                          & 75.12                        & -0.5                 & 5.02                     \\
\textbf{T1100-D2}     & 2                          & 34.93                        & 0.94                 & 0.94  \\\bottomrule
\textbf{Average}    & 5 & 51.28 & 4.2 & 7.05 \\ \bottomrule 
\end{tabular}}
\caption{LDDT improvement over extreme challenging sequences(depth<10), MSAs selected by highest pLDDT and by highest LDDT}
\label{tab:criteria_detail}
\end{table}

\begin{table*}[!htb]
\resizebox{\linewidth}{!}{

\begin{tabular}{cccccccccccccc}
\toprule
\multirow{2}{*}{\textbf{ID}} &\multirow{2}{*}{\textbf{Depth}} & \multicolumn{6}{c}{\textbf{pLDDT}} & \multicolumn{6}{c}{\textbf{LDDT}} \\ 
\cmidrule(lr){3-8}\cmidrule(r){9-14}

                  &                  & Org & Aug1 & Aug2 & Aug3 & Aug-EN        & $\Delta $pLDDT & Org & Aug1 & Aug2 & Aug3 & Aug-EN         & $\Delta $LDDT  \\ \hline
\textbf{T1026-D1}     & 20                           & 88.59 & 88.59 & 88.59 & 88.59 & 88.59 & 0           & 81.04 & 81.04 & 81.04 & 81.04 & 81.04 & 0          \\
\textbf{T1030-D2}     & 27                           & 88.03 & 88.03 & 88.03 & 88.03 & 88.03 & 0           & 83.11 & 83.11 & 83.11 & 83.11 & 83.11 & 0          \\
\textbf{T1037-D1}     & 4                            & 36.04 & 40.68 & 36.04 & 36.04 & 40.68 & 4.64        & 24.09 & 26.08 & 24.09 & 24.09 & 26.08 & 1.99       \\
\textbf{T1038-D1}     & 36                           & 79.09 & 79.09 & 79.09 & 79.09 & 79.09 & 0           & 73.67 & 73.67 & 73.67 & 73.67 & 73.67 & 0          \\
\textbf{T1038-D2}     & 33                           & 85.77 & 86.12 & 87.48 & 85.77 & 87.48 & 1.71        & 78.94 & 80.15 & 80.2  & 78.94 & 80.2  & 1.26       \\
\textbf{T1041-D1}     & 22                           & 80.25 & 80.25 & 80.15 & 80.25 & 80.25 & 0           & 62.95 & 62.95 & 53.35 & 62.95 & 62.95 & 0          \\
\textbf{T1042-D1}     & 2                            & 41.99 & 42.25 & 45.45 & 47.86 & 47.86 & 5.87        & 32.42 & 32.19 & 32.43 & 32.84 & 32.84 & 0.42       \\
\textbf{T1046s1-D1}   & 42                           & 93.92 & 94.59 & 94.84 & 94.97 & 94.97 & 1.05        & 88.51 & 88.16 & 88.04 & 88.77 & 88.77 & 0.26       \\
\textbf{T1046s2-D1}   & 44                           & 93.16 & 93.16 & 93.16 & 93.16 & 93.16 & 0           & 89.35 & 89.35 & 89.35 & 89.35 & 89.35 & 0          \\
\textbf{T1060s2-D1}   & 31                           & 82.09 & 85.19 & 85.91 & 87.64 & 87.64 & 5.55        & 77.18 & 75.62 & 77.36 & 72.59 & 72.59 & -4.59      \\
\textbf{T1064-D1}     & 9                            & 60.34 & 66.93 & 63.1  & 65    & 66.93 & 6.59        & 31.25 & 37.48 & 31.34 & 51.36 & 37.48 & 6.23       \\
\textbf{T1068-D1}     & 34                           & 92.38 & 93.76 & 92.38 & 92.93 & 92.93 & 0.55        & 90.8  & 89.95 & 90.8  & 90.03 & 90.03 & -0.77      \\
\textbf{T1074-D1}     & 9                            & 85.96 & 85.96 & 85.96 & 85.96 & 85.96 & 0           & 81.14 & 81.14 & 81.14 & 81.14 & 81.14 & 0          \\
\textbf{T1082-D1}     & 10                           & 92.31 & 92.79 & 92.31 & 92.31 & 92.79 & 0.48        & 87.37 & 88.49 & 87.37 & 87.37 & 88.49 & 1.12       \\
\textbf{T1093-D1}     & 3                            & 62.1  & 81.26 & 62.1  & 62.1  & 81.26 & 19.16       & 45.5  & 70.77 & 45.5  & 45.5  & 70.77 & 25.27      \\
\textbf{T1093-D3}     & 2                            & 74.11 & 74.11 & 74.11 & 74.11 & 74.11 & 0           & 66.5  & 66.5  & 66.5  & 66.5  & 66.5  & 0          \\
\textbf{T1094-D2}     & 7                            & 90.54 & 90.54 & 93.34 & 91.47 & 93.34 & 2.8         & 77.12 & 77.12 & 77.02 & 76.47 & 77.02 & -0.1       \\
\textbf{T1096-D1}     & 7                            & 70.28 & 73.06 & 86.25 & 83.28 & 86.25 & 15.97       & 61.92 & 62.83 & 71.19 & 69.23 & 71.19 & 9.27       \\
\textbf{T1096-D2}     & 2                            & 45.55 & 50.01 & 54.55 & 50.24 & 54.55 & 9           & 34.07 & 36.56 & 36.98 & 34.66 & 36.98 & 2.91       \\
\textbf{T1099-D1}     & 8                            & 89.27 & 89.99 & 89.27 & 90.38 & 90.38 & 1.11        & 75.12 & 74.27 & 75.12 & 74.62 & 74.62 & -0.5       \\
\textbf{T1100-D2}     & 2                            & 42.16 & 47.27 & 42.16 & 42.16 & 47.27 & 5.11        & 34.93 & 35.87 & 34.93 & 34.93 & 35.87 & 0.94       \\
\textbf{Average}      & {\color[HTML]{333333} 16.86} & 74.95 & 77.32 & 76.87 & 76.73 & 78.74 & 3.79        & 65.57 & 67.30 & 65.74 & 66.63 & 67.65 & 2.08      \\ \bottomrule
\end{tabular}

}
\caption{pLDDT (left) and LDDT (right) improvement over difficult MSA (depth<50).}
\label{tab:all_results}
\end{table*}

\subsubsection{Artificial Extreme Challenging Results}\label{sec:artificial results}
Results of all 81 artificial extreme challenging MSAs are shown in table \ref{tab:artificial results1} and table \ref{tab:artificial results2}.
\begin{table*}[ht]
\centering
\resizebox{0.8\textwidth}{!}{
\begin{tabular}{lcccccccc}
\toprule
MSA-ID        & \multicolumn{1}{l}{Gold} & \multicolumn{1}{l}{Original} & \multicolumn{1}{l}{Aug1} & \multicolumn{1}{l}{Aug2} & \multicolumn{1}{l}{Aug3} & \multicolumn{1}{l}{ENs} & \multicolumn{1}{l}{over original} & \multicolumn{1}{l|}{over gold} \\ \hline
T1024-D1   & 87.10                    & 81.56                        & 77.25                    & 81.62                    & 81.62                    & 81.62                   & 0.06                                       & -5.48                                  \\
T1024-D2   & 84.44                    & 86.66                        & 79.67                    & 82.84                    & 82.84                    & 82.84                   & -3.82                                      & -1.60                                  \\
T1025-D1   & 83.81                    & 82.04                        & 78.92                    & 76.61                    & 76.61                    & 78.92                   & -3.12                                      & -4.89                                  \\
T1026-D1   & 82.27                    & 48.72                        & 22.59                    & 79.82                    & 79.82                    & 79.82                   & 31.10                                      & -2.45                                  \\
T1027-D1   & 48.81                    & 39.15                        & 37.31                    & 46.38                    & 46.38                    & 46.38                   & 7.23                                       & -2.43                                  \\
T1028-D1   & 75.57                    & 32.35                        & 52.62                    & 52.09                    & 52.09                    & 52.62                   & 20.27                                      & -22.95                                 \\
T1029-D1   & 47.62                    & 47.43                        & 47.76                    & 47.44                    & 47.44                    & 47.76                   & 0.33                                       & 0.14                                   \\
T1030-D1   & 86.85                    & 64.92                        & 86.68                    & 86.56                    & 86.56                    & 86.68                   & 21.76                                      & -0.17                                  \\
T1030-D2   & 82.25                    & 80.99                        & 81.76                    & 82.06                    & 82.06                    & 82.06                   & 1.07                                       & -0.19                                  \\
T1031-D1   & 65.65                    & 43.58                        & 30.75                    & 37.99                    & 37.99                    & 37.99                   & -5.59                                      & -27.66                                 \\
T1032-D1   & 22.48                    & 19.90                        & 68.50                    & 62.11                    & 62.11                    & 68.50                   & 48.60                                      & 46.02                                  \\
T1034-D1   & 86.75                    & 83.15                        & 83.65                    & 84.05                    & 84.05                    & 84.05                   & 0.90                                       & -2.70                                  \\
T1035-D1   & 32.52                    & 33.78                        & 34.56                    & 36.32                    & 36.32                    & 36.32                   & 2.54                                       & 3.80                                   \\
T1036s1-D1 & 67.02                    & 26.11                        & 79.99                    & 79.13                    & 79.13                    & 79.99                   & 53.88                                      & 12.97                                  \\
T1038-D1   & 55.80                    & 26.03                        & 38.04                    & 27.61                    & 27.61                    & 38.04                   & 12.01                                      & -17.76                                 \\
T1038-D2   & 80.60                    & 76.37                        & 61.02                    & 58.57                    & 53.99                    & 61.02                   & -15.35                                     & -19.58                                 \\
T1041-D1   & 62.13                    & 45.60                        & 32.14                    & 0.00                     & 34.74                    & 34.74                   & -10.86                                     & -27.39                                 \\
T1045s1-D1 & 56.74                    & 33.67                        & 89.22                    & 83.99                    & 88.47                    & 89.22                   & 55.55                                      & 32.48                                  \\
T1045s2-D1 & 27.65                    & 24.44                        & 22.81                    & 24.89                    & 21.61                    & 24.89                   & 0.45                                       & -2.76                                  \\
T1046s1-D1 & 87.67                    & 72.65                        & 87.33                    & 0.00                     & 87.98                    & 87.98                   & 15.33                                      & 0.31                                   \\
T1046s2-D1 & 63.88                    & 25.42                        & 51.58                    & 52.44                    & 36.05                    & 52.44                   & 27.02                                      & -11.44                                 \\
T1047s1-D1 & 58.76                    & 60.01                        & 59.75                    & 61.65                    & 61.23                    & 61.65                   & 1.64                                       & 2.89                                   \\
T1047s2-D1 & 40.65                    & 39.12                        & 69.07                    & 70.64                    & 71.69                    & 71.69                   & 32.57                                      & 31.04                                  \\
T1047s2-D2 & 58.59                    & 59.58                        & 64.60                    & 64.30                    & 67.72                    & 67.72                   & 8.14                                       & 9.13                                   \\
T1047s2-D3 & 58.12                    & 51.61                        & 57.63                    & 58.18                    & 57.53                    & 58.18                   & 6.57                                       & 0.06                                   \\
T1048-D1   & 70.60                    & 72.34                        & 71.81                    & 71.75                    & 71.82                    & 71.82                   & -0.52                                      & 1.22                                   \\
T1049-D1   & 72.85                    & 33.01                        & 80.28                    & 81.03                    & 72.12                    & 81.03                   & 48.02                                      & 8.18                                   \\
T1050-D1   & 89.66                    & 88.46                        & 73.27                    & 72.51                    & 72.97                    & 73.27                   & -15.19                                     & -16.39                                 \\
T1050-D2   & 81.05                    & 76.84                        & 76.81                    & 78.93                    & 0.00                     & 78.93                   & 2.09                                       & -2.12                                  \\
T1050-D3   & 71.11                    & 70.25                        & 48.28                    & 51.18                    & 49.74                    & 51.18                   & -19.07                                     & -19.93                                 \\
T1052-D1   & 77.26                    & 68.84                        & 82.18                    & 83.82                    & 83.08                    & 83.82                   & 14.98                                      & 6.56                                   \\
T1052-D2  & 48.22                    & 34.95                        & 57.37                    & 59.27                    & 30.91                    & 59.27                   & 24.32                                      & 11.05                                  \\
T1052-D3   & 90.81                    & 82.12                        & 89.62                    & 85.22                    & 0.00                     & 89.62                   & 7.50                                       & -1.19                                  \\
T1053-D1   & 72.86                    & 30.58                        & 73.75                    & 74.09                    & 65.32                    & 74.09                   & 43.51                                      & 1.23                                   \\
T1053-D2   & 75.75                    & 75.61                        & 77.31                    & 76.25                    & 0.00                     & 77.31                   & 1.70                                       & 1.56                                   \\
T1054-D1   & 34.64                    & 31.68                        & 70.68                    & 81.44                    & 0.00                     & 81.44                   & 49.76                                      & 46.80                                  \\
T1055-D1   & 67.90                    & 55.88                        & 19.03                    & 25.46                    & 21.62                    & 25.46                   & -30.42                                     & -42.44                                 \\ \bottomrule
                               
\end{tabular}}
\caption{LDDT improvement over artificial extremely challenging dataset (First half). Follow the same notations as in \ref{sec:notation}}
\label{tab:artificial results1}
\end{table*}

\begin{table*}[ht]
\centering
\resizebox{0.8\textwidth}{!}{
\begin{tabular}{ccccccccc}\toprule
MSA-ID        & \multicolumn{1}{l}{Gold} & \multicolumn{1}{l}{Original} & \multicolumn{1}{l}{Aug1} & \multicolumn{1}{l}{Aug2} & \multicolumn{1}{l}{Aug3} & \multicolumn{1}{l}{ENs} & \multicolumn{1}{l}{over original} & \multicolumn{1}{l|}{over gold} \\ \hline
T1056-D1   & 75.61                    & 76.11                        & 64.62                    & 65.39                    & 62.93                    & 65.39                   & -10.72                                     & -10.22                                 \\
T1057-D1   & 86.18                    & 52.96                        & 81.38                    & 80.30                    & 81.04                    & 81.38                   & 28.42                                      & -4.80                                  \\
T1058-D1   & 45.72                    & 38.10                        & 37.47                    & 41.31                    & 39.76                    & 41.31                   & 3.21                                       & -4.41                                  \\
T1058-D2   & 51.18                    & 29.59                        & 61.13                    & 60.41                    & 60.73                    & 61.13                   & 31.54                                      & 9.95                                   \\
T1060s2-D1 & 75.25                    & 38.44                        & 72.84                    & 74.01                    & 0.00                     & 74.01                   & 35.57                                      & -1.24                                  \\
T1060s3-D1 & 80.69                    & 69.93                        & 78.72                    & 78.31                    & 78.89                    & 78.89                   & 8.96                                       & -1.80                                  \\
T1061-D0   & 60.24                    & 57.51                        & 54.74                    & 0.00                     & 58.36                    & 58.36                   & 0.85                                       & 0.85                                   \\
T1061-D1   & 37.50                    & 24.28                        & 51.45                    & 46.94                    & 50.60                    & 51.45                   & 27.17                                      & 13.95                                  \\
T1061-D2   & 53.34                    & 31.82                        & 58.05                    & 50.32                    & 55.92                    & 58.05                   & 26.23                                      & 4.71                                   \\
T1061-D3   & 81.92                    & 42.92                        & 77.44                    & 77.34                    & 76.29                    & 77.44                   & 34.52                                      & -4.48                                  \\
T1062-D1   & 59.10                    & 72.66                        & 72.61                    & 59.58                    & 0.00                     & 72.61                   & -0.05                                      & 13.51                                  \\
T1065s1-D1 & 88.59                    & 44.32                        & 90.31                    & 91.00                    & 0.00                     & 91.00                   & 46.68                                      & 2.41                                   \\
T1065s2-D1 & 70.46                    & 63.90                        & 80.77                    & 87.09                    & 83.49                    & 87.09                   & 23.19                                      & 16.63                                  \\
T1067-D1   & 78.55                    & 27.33                        & 79.05                    & 75.29                    & 77.91                    & 79.05                   & 51.72                                      & 0.50                                   \\
T1068-D1   & 85.05                    & 35.20                        & 89.55                    & 88.93                    & 89.92                    & 89.92                   & 54.72                                      & 4.87                                   \\
T1070-D1   & 53.44                    & 49.68                        & 55.69                    & 52.55                    & 52.73                    & 55.69                   & 6.01                                       & 2.25                                   \\
T1070-D2   & 29.16                    & 35.04                        & 87.50                    & 86.84                    & 90.38                    & 90.38                   & 55.34                                      & 61.22                                  \\
T1070-D3   & 74.12                    & 73.32                        & 73.56                    & 73.50                    & 0.00                     & 73.56                   & 0.24                                       & -0.56                                  \\
T1070-D4   & 73.83                    & 30.62                        & 74.39                    & 84.71                    & 85.07                    & 85.07                   & 54.45                                      & 11.24                                  \\
T1073-D1   & 76.17                    & 76.19                        & 76.79                    & 0.00                     & 76.11                    & 76.79                   & 0.60                                       & 0.62                                   \\
T1076-D1   & 83.10                    & 65.52                        & 71.31                    & 68.61                    & 68.35                    & 71.31                   & 5.79                                       & -11.79                                 \\
T1078-D1   & 77.22                    & 54.56                        & 0.00                     & 64.61                    & 72.25                    & 72.25                   & 17.69                                      & -4.97                                  \\
T1079-D1   & 84.51                    & 63.88                        & 79.44                    & 69.34                    & 68.08                    & 79.44                   & 15.56                                      & -5.07                                  \\
T1080-D1   & 63.68                    & 59.76                        & 67.91                    & 67.67                    & 68.08                    & 68.08                   & 8.32                                       & 4.40                                   \\
T1083-D1   & 78.49                    & 78.26                        & 78.02                    & 77.48                    & 77.72                    & 78.02                   & -0.24                                      & -0.47                                  \\
T1084-D1   & 86.64                    & 86.47                        & 49.09                    & 86.03                    & 86.25                    & 86.25                   & -0.22                                      & -0.39                                  \\
T1087-D1   & 80.76                    & 77.32                        & 37.14                    & 39.73                    & 69.27                    & 69.27                   & -8.05                                      & -11.49                                 \\
T1088-D1   & 26.48                    & 24.00                        & 33.92                    & 34.10                    & 54.40                    & 54.40                   & 30.40                                      & 27.92                                  \\
T1089-D1   & 83.99                    & 77.55                        & 64.70                    & 64.30                    & 67.15                    & 67.15                   & -10.40                                     & -16.84                                 \\
T1090-D1   & 76.73                    & 55.34                        & 57.65                    & 0.00                     & 56.06                    & 57.65                   & 2.31                                       & -19.08                                 \\
T1091-D1   & 43.89                    & 43.26                        & 70.35                    & 71.45                    & 76.76                    & 76.76                   & 33.50                                      & 32.87                                  \\
T1091-D2   & 81.76                    & 40.56                        & 50.57                    & 0.00                     & 47.30                    & 50.57                   & 10.01                                      & -31.19                                 \\
T1091-D3   & 71.58                    & 52.00                        & 75.93                    & 0.00                     & 75.98                    & 75.98                   & 23.98                                      & 4.40                                   \\
T1091-D4   & 79.76                    & 56.64                        & 83.20                    & 0.00                     & 78.01                    & 83.20                   & 26.56                                      & 3.44                                   \\
T1092-D1   & 63.77                    & 56.56                        & 33.97                    & 38.95                    & 40.48                    & 40.48                   & -16.08                                     & -23.29                                 \\
T1092-D2   & 70.61                    & 53.20                        & 43.76                    & 26.54                    & 43.80                    & 43.80                   & -9.40                                      & -26.81                                 \\
T1093-D2   & 29.91                    & 34.69                        & 28.20                    & 28.12                    & 31.08                    & 31.08                   & -3.61                                      & 1.17                                   \\
T1094-D1   & 28.74                    & 26.73                        & 21.89                    & 21.63                    & 20.72                    & 21.89                   & -4.84                                      & -6.85                                  \\
T1095-D1   & 60.06                    & 55.57                        & 35.95                    & 36.50                    & 32.78                    & 36.50                   & -19.07                                     & -23.56                                 \\
T1098-D1   & 57.45                    & 29.82                        & 49.36                    & 45.94                    & 36.76                    & 49.36                   & 19.54                                      & -8.09                                  \\
T1098-D2   & 37.88                    & 31.58                        & 45.70                    & 45.89                    & 45.36                    & 45.89                   & 14.31                                      & 8.01                                   \\
T1100-D1   & 57.78                    & 65.70                        & 60.39                    & 60.01                    & 61.10                    & 61.10                   & -4.60                                      & 3.32                                   \\
T1101-D1   & 88.14                    & 87.57                        & 87.95                    & 87.64                    & 86.52                    & 87.95                   & 0.38                                       & -0.19                                  \\
T1101-D2   & 78.74                    & 77.81                        & 43.27                    & 42.09                    & 74.91                    & 74.91                   & -2.90                                      & -3.83                                  \\ \bottomrule
Average    & 66.47                    & 53.45                        & 61.77                    & 57.14                    & 56.43                    & 66.32                   & 12.87                                      & -0.11  
              \\\bottomrule                 
\end{tabular}}
\caption{LDDT improvement over artificial extremely challenging dataset (Second half). Follow the same notations as in \ref{sec:notation}}
\label{tab:artificial results2}
\end{table*}

%% file: neurips_2023.bbl
\begin{thebibliography}{10}

\bibitem{Beltagy2020LongformerTL}
Iz~Beltagy, Matthew~E. Peters, and Arman Cohan.
\newblock Longformer: The long-document transformer.
\newblock {\em ArXiv}, abs/2004.05150, 2020.

\bibitem{codex}
Mark Chen, Jerry Tworek, Heewoo Jun, Qiming Yuan, Henrique~Ponde
  de~Oliveira~Pinto, Jared Kaplan, Harri Edwards, Yuri Burda, Nicholas Joseph,
  Greg Brockman, Alex Ray, Raul Puri, Gretchen Krueger, Michael Petrov, Heidy
  Khlaaf, Girish Sastry, Pamela Mishkin, Brooke Chan, Scott Gray, Nick Ryder,
  Mikhail Pavlov, Alethea Power, Lukasz Kaiser, Mohammad Bavarian, Clemens
  Winter, Philippe Tillet, Felipe~Petroski Such, Dave Cummings, Matthias
  Plappert, Fotios Chantzis, Elizabeth Barnes, Ariel Herbert-Voss,
  William~Hebgen Guss, Alex Nichol, Alex Paino, Nikolas Tezak, Jie Tang, Igor
  Babuschkin, Suchir Balaji, Shantanu Jain, William Saunders, Christopher
  Hesse, Andrew~N. Carr, Jan Leike, Josh Achiam, Vedant Misra, Evan Morikawa,
  Alec Radford, Matthew Knight, Miles Brundage, Mira Murati, Katie Mayer, Peter
  Welinder, Bob McGrew, Dario Amodei, Sam McCandlish, Ilya Sutskever, and
  Wojciech Zaremba.
\newblock Evaluating large language models trained on code, 2021.

\bibitem{chowdhury2022single}
Ratul Chowdhury, Nazim Bouatta, Surojit Biswas, Christina Floristean, Anant
  Kharkar, Koushik Roy, Charlotte Rochereau, Gustaf Ahdritz, Joanna Zhang,
  George~M Church, et~al.
\newblock Single-sequence protein structure prediction using a language model
  and deep learning.
\newblock {\em Nature Biotechnology}, 40(11):1617--1623, 2022.

\bibitem{chowdhury2021single}
Ratul Chowdhury, Nazim Bouatta, Surojit Biswas, Charlotte Rochereau, George~M
  Church, Peter~Karl Sorger, and Mohammed~N AlQuraishi.
\newblock Single-sequence protein structure prediction using language models
  from deep learning.
\newblock {\em bioRxiv}, 2021.

\bibitem{flant5}
Hyung~Won Chung, Le~Hou, Shayne Longpre, Barret Zoph, Yi~Tay, William Fedus,
  Yunxuan Li, Xuezhi Wang, Mostafa Dehghani, Siddhartha Brahma, Albert Webson,
  Shixiang~Shane Gu, Zhuyun Dai, Mirac Suzgun, Xinyun Chen, Aakanksha
  Chowdhery, Alex Castro-Ros, Marie Pellat, Kevin Robinson, Dasha Valter,
  Sharan Narang, Gaurav Mishra, Adams Yu, Vincent Zhao, Yanping Huang, Andrew
  Dai, Hongkun Yu, Slav Petrov, Ed~H. Chi, Jeff Dean, Jacob Devlin, Adam
  Roberts, Denny Zhou, Quoc~V. Le, and Jason Wei.
\newblock Scaling instruction-finetuned language models, 2022.

\bibitem{Dai2020FunnelTransformerFO}
Zihang Dai, Guokun Lai, Yiming Yang, and Quoc~V. Le.
\newblock Funnel-transformer: Filtering out sequential redundancy for efficient
  language processing.
\newblock {\em ArXiv}, abs/2006.03236, 2020.

\bibitem{fedus2021switch}
William Fedus, Barret Zoph, and Noam Shazeer.
\newblock Switch transformers: Scaling to trillion parameter models with simple
  and efficient sparsity, 2021.

\bibitem{guo2020bagging}
Yuzhi Guo, Jiaxiang Wu, Hehuan Ma, Sheng Wang, and Junzhou Huang.
\newblock Bagging msa learning: Enhancing low-quality pssm with deep learning
  for accurate protein structure property prediction.
\newblock In {\em International Conference on Research in Computational
  Molecular Biology}, pages 88--103. Springer, 2020.

\bibitem{ho2019axial}
Jonathan Ho, Nal Kalchbrenner, Dirk Weissenborn, and Tim Salimans.
\newblock Axial attention in multidimensional transformers.
\newblock {\em arXiv preprint arXiv:1912.12180}, 2019.

\bibitem{johnson2010hidden}
L~Steven Johnson, Sean~R Eddy, and Elon Portugaly.
\newblock Hidden markov model speed heuristic and iterative hmm search
  procedure.
\newblock {\em BMC bioinformatics}, 11(1):1--8, 2010.

\bibitem{jumper2021highly}
John Jumper, Richard Evans, Alexander Pritzel, Tim Green, Michael Figurnov,
  Olaf Ronneberger, Kathryn Tunyasuvunakool, Russ Bates, Augustin
  {\v{Z}}{\'\i}dek, Anna Potapenko, et~al.
\newblock Highly accurate protein structure prediction with alphafold.
\newblock {\em Nature}, 596(7873):583--589, 2021.

\bibitem{Katharopoulos2020TransformersAR}
Angelos Katharopoulos, Apoorv Vyas, Nikolaos Pappas, and Franccois Fleuret.
\newblock Transformers are rnns: Fast autoregressive transformers with linear
  attention.
\newblock In {\em International Conference on Machine Learning}, 2020.

\bibitem{lewis2019bart}
Mike Lewis, Yinhan Liu, Naman Goyal, Marjan Ghazvininejad, Abdelrahman Mohamed,
  Omer Levy, Ves Stoyanov, and Luke Zettlemoyer.
\newblock Bart: Denoising sequence-to-sequence pre-training for natural
  language generation, translation, and comprehension.
\newblock {\em arXiv preprint arXiv:1910.13461}, 2019.

\bibitem{lin2022language}
Zeming Lin, Halil Akin, Roshan Rao, Brian Hie, Zhongkai Zhu, Wenting Lu, Nikita
  Smetanin, Allan dos Santos~Costa, Maryam Fazel-Zarandi, Tom Sercu, Sal
  Candido, et~al.
\newblock Language models of protein sequences at the scale of evolution enable
  accurate structure prediction.
\newblock {\em bioRxiv}, 2022.

\bibitem{madani2020progen}
Ali Madani, Bryan McCann, Nikhil Naik, Nitish~Shirish Keskar, Namrata Anand,
  Raphael~R Eguchi, Po-Ssu Huang, and Richard Socher.
\newblock Progen: Language modeling for protein generation.
\newblock {\em arXiv preprint arXiv:2004.03497}, 2020.

\bibitem{mariani2013lddt}
Valerio Mariani, Marco Biasini, Alessandro Barbato, and Torsten Schwede.
\newblock lddt: a local superposition-free score for comparing protein
  structures and models using distance difference tests.
\newblock {\em Bioinformatics}, 29(21):2722--2728, 2013.

\bibitem{mirdita2017uniclust}
Milot Mirdita, Lars Von Den~Driesch, Clovis Galiez, Maria~J Martin, Johannes
  S{\"o}ding, and Martin Steinegger.
\newblock Uniclust databases of clustered and deeply annotated protein
  sequences and alignments.
\newblock {\em Nucleic acids research}, 45(D1):D170--D176, 2017.

\bibitem{nijkamp2022progen2}
Erik Nijkamp, Jeffrey Ruffolo, Eli~N Weinstein, Nikhil Naik, and Ali Madani.
\newblock Progen2: Exploring the boundaries of protein language models.”
  arxiv, 2022.

\bibitem{raffel2020exploring}
Colin Raffel, Noam Shazeer, Adam Roberts, Katherine Lee, Sharan Narang, Michael
  Matena, Yanqi Zhou, Wei Li, Peter~J Liu, et~al.
\newblock Exploring the limits of transfer learning with a unified text-to-text
  transformer.
\newblock {\em J. Mach. Learn. Res.}, 21(140):1--67, 2020.

\bibitem{rao2020transformer}
Roshan Rao, Joshua Meier, Tom Sercu, Sergey Ovchinnikov, and Alexander Rives.
\newblock Transformer protein language models are unsupervised structure
  learners.
\newblock {\em Biorxiv}, 2020.

\bibitem{pmlr-v139-rao21a}
Roshan~M Rao, Jason Liu, Robert Verkuil, Joshua Meier, John Canny, Pieter
  Abbeel, Tom Sercu, and Alexander Rives.
\newblock Msa transformer.
\newblock In Marina Meila and Tong Zhang, editors, {\em Proceedings of the 38th
  International Conference on Machine Learning}, volume 139 of {\em Proceedings
  of Machine Learning Research}, pages 8844--8856. PMLR, 18--24 Jul 2021.

\bibitem{rives2021biological}
Alexander Rives, Joshua Meier, Tom Sercu, Siddharth Goyal, Zeming Lin, Jason
  Liu, Demi Guo, Myle Ott, C~Lawrence Zitnick, Jerry Ma, et~al.
\newblock Biological structure and function emerge from scaling unsupervised
  learning to 250 million protein sequences.
\newblock {\em Proceedings of the National Academy of Sciences},
  118(15):e2016239118, 2021.

\bibitem{sgarbossa2022generative}
Damiano Sgarbossa, Umberto Lupo, and Anne-Florence Bitbol.
\newblock Generative power of a protein language model trained on multiple
  sequence alignments.
\newblock {\em bioRxiv}, 2022.

\bibitem{suzek2007uniref}
Baris~E Suzek, Hongzhan Huang, Peter McGarvey, Raja Mazumder, and Cathy~H Wu.
\newblock Uniref: comprehensive and non-redundant uniprot reference clusters.
\newblock {\em Bioinformatics}, 23(10):1282--1288, 2007.

\bibitem{touvron2023llama}
Hugo Touvron, Thibaut Lavril, Gautier Izacard, Xavier Martinet, Marie-Anne
  Lachaux, Timothée Lacroix, Baptiste Rozière, Naman Goyal, Eric Hambro,
  Faisal Azhar, Aurelien Rodriguez, Armand Joulin, Edouard Grave, and Guillaume
  Lample.
\newblock Llama: Open and efficient foundation language models, 2023.

\bibitem{vaswani2017attention}
Ashish Vaswani, Noam Shazeer, Niki Parmar, Jakob Uszkoreit, Llion Jones,
  Aidan~N Gomez, {\L}ukasz Kaiser, and Illia Polosukhin.
\newblock Attention is all you need.
\newblock {\em Advances in neural information processing systems}, 30, 2017.

\bibitem{vig2020bertology}
Jesse Vig, Ali Madani, Lav~R Varshney, Caiming Xiong, Richard Socher, and
  Nazneen~Fatema Rajani.
\newblock Bertology meets biology: interpreting attention in protein language
  models.
\newblock {\em arXiv preprint arXiv:2006.15222}, 2020.

\bibitem{wang2022contact}
Qin Wang, Jiayang Chen, Yuzhe Zhou, Yu~Li, Liangzhen Zheng, Sheng Wang, Zhen
  Li, and Shuguang Cui.
\newblock Contact-distil: Boosting low homologous protein contact map
  prediction by self-supervised distillation.
\newblock In {\em Proceedings of the AAAI Conference on Artificial
  Intelligence}, volume~36, pages 4620--4627, 2022.

\bibitem{wang2021pssm}
Qin Wang, Boyuan Wang, Zhenlei Xu, Jiaxiang Wu, Peilin Zhao, Zhen Li, Sheng
  Wang, Junzhou Huang, and Shuguang Cui.
\newblock Pssm-distil: Protein secondary structure prediction (pssp) on
  low-quality pssm by knowledge distillation with contrastive learning.
\newblock In {\em Proceedings of the AAAI Conference on Artificial
  Intelligence}, volume~35, pages 617--625, 2021.

\bibitem{wu2022high}
Ruidong Wu, Fan Ding, Rui Wang, Rui Shen, Xiwen Zhang, Shitong Luo, Chenpeng
  Su, Zuofan Wu, Qi~Xie, Bonnie Berger, et~al.
\newblock High-resolution de novo structure prediction from primary sequence.
\newblock {\em BioRxiv}, 2022.

\bibitem{Zaheer2020BigBT}
Manzil Zaheer, Guru Guruganesh, Kumar~Avinava Dubey, Joshua Ainslie, Chris
  Alberti, Santiago Onta{\~n}{\'o}n, Philip Pham, Anirudh Ravula, Qifan Wang,
  Li~Yang, and Amr Ahmed.
\newblock Big bird: Transformers for longer sequences.
\newblock {\em ArXiv}, abs/2007.14062, 2020.

\end{thebibliography}
